\documentclass[aip, reprint]{revtex4-2}
\usepackage{braket}
\usepackage{graphicx}
\usepackage{dcolumn}
\usepackage{bm}
\usepackage{textcmds}
\usepackage[utf8]{inputenc}
\usepackage[T1]{fontenc}
\usepackage{mathptmx}
\usepackage{xcolor}
\usepackage{etoolbox}
\usepackage[hidelinks]{hyperref}
\usepackage[nameinlink]{cleveref}
\makeatletter
\def\@email#1#2{%
 \endgroup
 \patchcmd{\titleblock@produce}
  {\frontmatter@RRAPformat}
  {\frontmatter@RRAPformat{\produce@RRAP{*#1\href{mailto:#2}{#2}}}\frontmatter@RRAPformat}
  {}{}
}%
\makeatother
\begin{document}

\preprint{AIP/123-QED}
\title{Generation of Correlated Quantum Random Number Sequences with Bright Twin Beams}
\author{Anirudh Shekar}
\author{Chirang R. Patel}
\author{Jerin A. Thachil}
\author{Ashok Kumar$^*$}
 \email{ashokkumar@iist.ac.in}
\affiliation{Department of Physics, Indian Institute of Space Science and Technology, Thiruvananthapuram, Kerala, India - 695547.
}%

\date{\today}

\begin{abstract}
Quantum random number generators play a vital role in securing communication and encryption. In the present work, we have produced bright twin beams using four-wave mixing in a double-$\Lambda$ configuration in rubidium-85 vapor and generated two strings of highly correlated random numbers. The randomness originates from the probabilistic nature of the intensity fluctuations of the twin beams and the quantum correlations are certified by measuring the intensity-difference squeezing in the generated twin beams. At an analysis frequency of 2~MHz, we have measured 95\% correlation between the random intensity fluctuations of the twin beams. We observe over 5 bits/sample of entropy from the quantum fluctuations of the twin beams. Furthermore, to extract identical strings of random numbers, post-selection of the binned data and hashing algorithms are used, leading to a binary string of random numbers at a rate of 6~Mbps that passes standard statistical tests from NIST and TestU01. Here, the simplicity of generating bright twin beams shows the potential of this method in quantum cryptography and quantum communication.
\end{abstract}

\maketitle

A quantum random number generator (QRNG) produces strings of random numbers, where the randomness originates from the probabilistic nature of quantum mechanics\cite{herrero2017quantum,bera2017randomness,ma2016quantum}. Such a generator is different from the pseudo-random number generators (PRNGs) that are more widely used, where a deterministic algorithm produces binary strings based on the value of an input seed\cite{james1990review,knuth2014art}. The deterministic nature of these generators allows an adversary to compute the entire string if they can guess the input seed. The input seed is usually derived from chaotic processes that are computationally intensive to characterize, but in principle can be done with computational advancements in the future\cite{herrero2017quantum}. Thus, QRNGs form an integral part of quantum key distribution (QKD) protocols\cite{herrero2017quantum}. 

Several methods are proposed to produce quantum random numbers\cite{herrero2017quantum,isida1956random,manelis1961generating,stipvcevic2004fast,vartsky2011high,jennewein2000fast,weihs1998violation,gabriel2010generator,shen2010practical,stipvcevic2007quantum,khanmohammadi2015monolithic,guo2010truly,qi2010high,furst2010high,soares2014quantum,zhang2017quantum} after the initial proposal based on the radioactive decay\cite{isida1956random,manelis1961generating}. In particular, the two most common methods are based on electronic and optical generation of the random numbers \cite{herrero2017quantum}. In the class of electronic QRNGs, noise in Zener diodes\cite{stipvcevic2004fast} and electron tunneling\cite{vartsky2011high} have been used to generate the random numbers. In the class of optical QRNGs, one can use a single photon incident on a 50:50 beam splitter as a QRNG by detecting which path the photon took after passing through the beam splitter\cite{jennewein2000fast,weihs1998violation}. Furthermore, the vacuum fluctuations\cite{gabriel2010generator,shen2010practical}, the time of arrival of photons\cite{stipvcevic2007quantum,khanmohammadi2015monolithic}, the phase noise of lasers\cite{guo2010truly,qi2010high} and photon counting\cite{furst2010high,soares2014quantum} have been used to generate random numbers.

In the present work, we optically generate two strings of random numbers that are highly correlated with each other, using the bright twin beams generated with a four-wave mixing\cite{mccormick2008strong} (FWM) process in a double-$\Lambda$ configuration in a hot rubidium atomic vapor cell. The use of FWM in our work enables us to obtain more than 95\% correlation between the intensity fluctuations of the bright twin beams obtained from the process. The quantumness of these correlations is certified by obtaining $\sim$6~dB of intensity-difference squeezing\cite{barnett1991information}. Such quantum correlations facilitate the extraction of identical strings of truly random numbers without significant post-processing.  One must note that, in contrast to the bright twin beams generated with optical parametric oscillators\cite{heidmann1987observation}, the FWM process in double-$\Lambda$ configuration allows one to generate bright twin beams without an optical cavity, thus making the system simpler and robust.

In what follows, we start with a theoretical description of how the photon number fluctuations in the bright twin beams or two-mode squeezed states lead to two strings of quantum-correlated random numbers. We then describe our experimental scheme for the generation of bright two-mode squeezed states, followed by the results and analysis of the generated quantum random numbers. 

\begin{figure*}[t]
    \centering
    \includegraphics[width=1\linewidth]{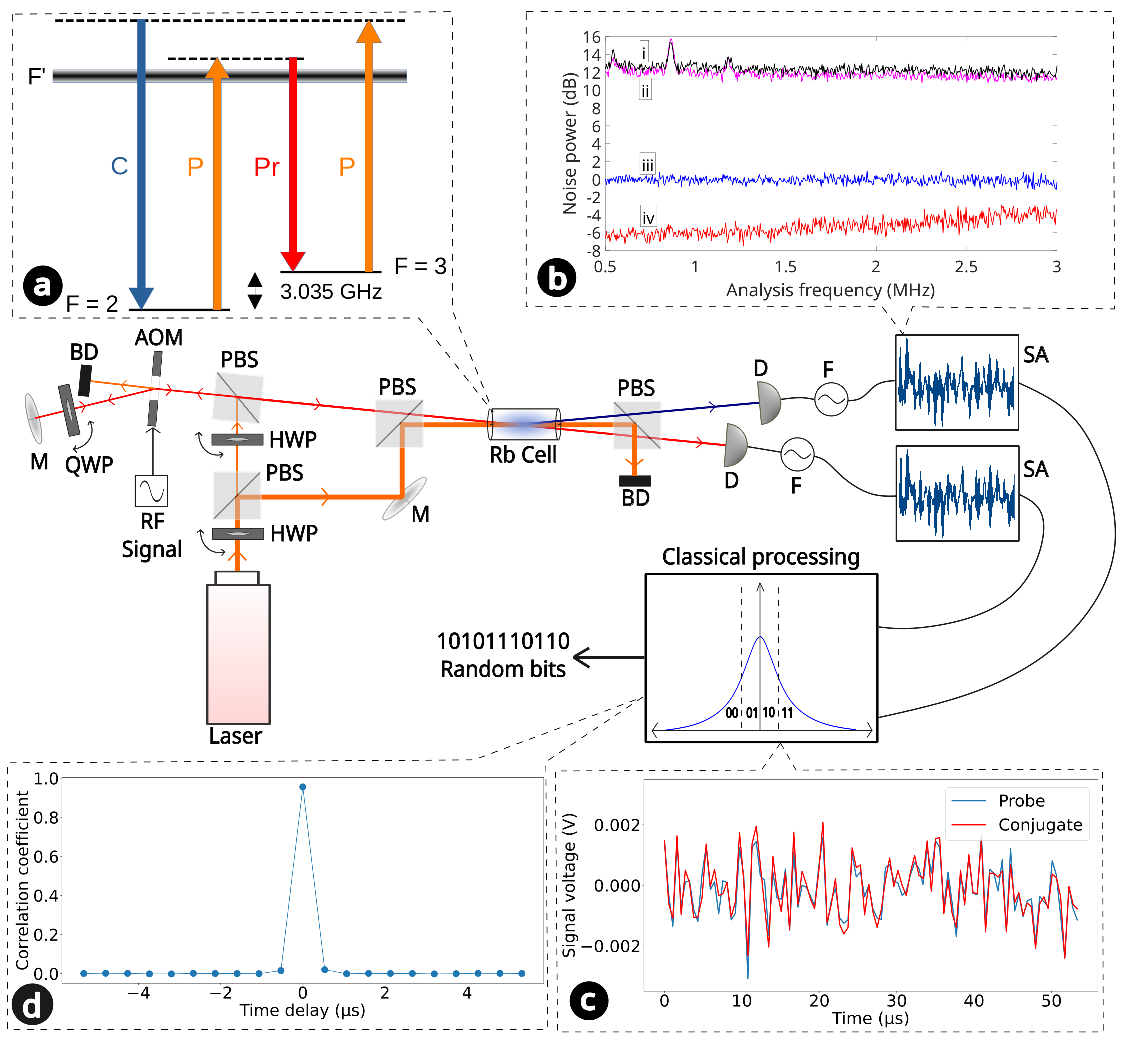}
    \caption{Schematic used to produce and extract the correlated quantum random numbers from the twin beams generated with a four-wave mixing process in a double-$\Lambda$ configuration (shown in inset (a)) in a hot rubidium-vapor cell. In the FWM process, two photons from the pump (P) are annihilated and two new photons, called the probe (Pr) and conjugate (C) are created. (b) Intensity noise power normalized by the shot noise as a function of analysis frequency, shown from top to bottom: individual conjugate noise (i), probe noise (ii), shot noise limit (iii), and intensity-difference noise of probe and conjugate beams (iv). (c) Correlated probe and conjugate intensity fluctuations with time. (d) Normalized correlation coefficients of the intensity fluctuations, as a function of time delays between the twin beams. We obtain a correlation coefficient of 0.957 with no delay. HWP: half wave plate; PBS: polarizing beam splitter; M: mirror; AOM: acousto-optic modulator; QWP: quarter wave plate; BD: beam dump; D: detector; F: high pass filter (RF bias tee); SA: spectrum analyzer.}
    \label{fig:1}
\end{figure*}
A two-mode squeezed state is given by 
\begin{equation}
    \ket{\alpha,\beta,\zeta} = \hat{S}(\zeta)\hat{D}_a(\alpha)\hat{D}_b(\beta)\ket{0,0},
    \label{tmss}
\end{equation}
where $\hat{S}(\zeta) = \text{exp}(\zeta^*\hat{a}\hat{b} -\zeta\hat{a}^\dagger\hat{b}^\dagger)$ is the two-mode squeezing operator\cite{caves1985new} with parameter $\zeta = se^{i\theta}$. Here, $s$ is the degree of squeezing and $\theta$ is the squeezing phase. $\hat{a}$ and $\hat{b}$ represent the annihilation operators for the two modes. $\hat{D}_a(\alpha)$ and $\hat{D}_b(\beta)$ are the displacement operators\cite{glauber1963coherent} with complex coherent amplitudes of $\alpha$ and $\beta$, respectively. 

Equation~\ref{tmss} is a general form of the two-mode squeezed states, from which the squeezed vacuum or the displaced squeezed vacuum states can be obtained by adjusting $|\alpha|$ and $|\beta|$. In the current work, we consider the case when $|\beta|=0$ and $|\alpha|^2\gg1$, which allows us to generate bright twin beams. In this limit, the average number of photons in each mode becomes
\begin{equation}
    \braket{\hat{n}_a} = |\alpha|^2\text{cosh}^2s = |\alpha|^2\text{G},
\end{equation}
\begin{equation}
    \braket{\hat{n}_b} = |\alpha|^2\text{sinh}^2s = |\alpha|^2(\text{G}-1),
\end{equation}
where $\text{G} = \text{cosh}^2s$ is defined as the gain of the process. Similarly, the variances of the photon number in the two modes can be calculated as
\begin{equation}
\braket{(\Delta\hat{n}_a)^2} = |\alpha|^2\text{cosh}^2s\ (\text{cosh}^2s + \text{sinh}^2s)\ +\ \text{sinh}^2s\ \text{cosh}^2s,
\end{equation}
\begin{equation}
    \braket{(\Delta\hat{n}_b)^2} = |\alpha|^2\text{sinh}^2s\ (\text{cosh}^2s + \text{sinh}^2s)\ +\ \text{sinh}^2s\ \text{cosh}^2s.
\end{equation}
We then obtain the photon number variance of the individual modes, normalized to that of a coherent state of the same power (called the shot noise), as
\begin{equation}
    R_a = \frac{\braket{(\Delta\hat{n}_a)^2}}{\braket{\hat{n}_a}} = \text{cosh}^2s + \text{sinh}^2s = 2\text{G} - 1,
    \label{prnoise}
\end{equation}
\begin{equation}
    R_b = \frac{\braket{(\Delta\hat{n}_b)^2}}{\braket{\hat{n}_b}} = \text{cosh}^2s + \text{sinh}^2s = 2\text{G} - 1.
    \label{conoise}
\end{equation}
Here, we have used the fact that the variance of the number of photons in a coherent state is equal to its mean. One can see from Eqs. (\ref{prnoise}) and (\ref{conoise}) that for G > 1, the variances of photon numbers of both modes are many times their mean values and hence, exhibit super-Poissonian statistics. 

Similarly, the photon number-difference variance of the two-mode squeezed states, normalized to that of a coherent state, can be obtained as
\begin{equation}
    R_{ab} = \frac{\braket{(\Delta(\hat{n}_a-\hat{n}_b))^2}}{\braket{\hat{n}_a}+\braket{\hat{n}_b}} = \frac{1}{\text{cosh}^2s\ +\ \text{sinh}^2s} = \frac{1}{2\text{G}-1}.
    \label{intdiff}
\end{equation}
Here, $\braket{\hat{n}_a}+\braket{\hat{n}_b}$ is the mean number of photons in a coherent state and is equivalent to the mean number of photons in the two modes.

On comparing Eq.~(\ref{intdiff}) with Eqs.~(\ref{prnoise}) and (\ref{conoise}), one can see that the intensity (photon number) difference noise of the twin modes is reduced by a factor of $1/(2\text{G}-1)^2$ from the individual mode intensity noise. 

Equation~(\ref{intdiff}) gives the intensity noise reduction in an ideal lossless case. In order to derive the same for the practical, lossy situations, one can use a beam splitter model for the losses in each mode, and can obtain the relative intensity difference squeezing as\cite{jasperse2011relative}
\begin{equation}
    R = 1 + \frac{2(G -1)(G(\eta_a-\eta_b)^2-\eta_b^2)}{G\eta_a+(G-1)\eta_b},
    \label{squeezing}
\end{equation}
where $1-\eta_a$ and $1-\eta_b$ are the losses encountered by the two modes.

As it will be shown in the following sections, for our system, the quantum origin of the intensity fluctuations can be certified if the individual intensity noise of both beams calculated from the fundamental principles of quantum mechanics (using Eq.~(\ref{squeezing})) is in agreement with what is observed experimentally.

In FIG. \ref{fig:1}, we show a schematic of the experimental setup used to generate bright two-mode squeezed states with four-wave mixing in a double-$\Lambda$ configuration (see FIG. \ref{fig:1}(a)) in a hot rubidium vapor cell. A bright pump beam of 550~mW power and 795~nm wavelength, obtained from a Titanium-Sapphire laser, is incident on the vapor cell heated to 107~$^\circ$C. An orthogonally polarized probe beam derived from the same laser, which acts as a seed for the process, is incident on the cell at an angle of 0.4$^\circ$ with respect to the pump beam. The frequency of this beam is shifted down $\sim$3~GHz from that of the pump beam using an acousto-optic modulator (AOM) driven by a radio frequency (RF) signal. The pump and probe have corresponding beam waists of 550~$\mu$m and 350~$\mu$m at the centre of the cell. After the FWM process, the probe beam gets amplified by a factor of the gain (G) of the process, and in the current experiment, G is set to be $\sim$11.5. At the same time, a new beam called the conjugate is created, which is correlated in all properties with the output probe beam. These twin beams are separated from the pump beam, using a polarizing beam splitter, and sent to two separate photodetectors (D). The DC output of the photodetectors is filtered out using an RF bias tee, and the AC part is sent to the two spectrum analyzers (SAs), where the fluctuations in their intensities in the RF regime are measured. In Fig.~\ref{fig:1}(b), the intensity fluctuations of the individual probe and conjugate beams, normalized with the shot noise, are plotted as a function of the analysis frequency (traces i and ii). We measured that the individual probe and conjugate intensity noises are $\sim$11.65~dB and $\sim$12.23~dB above the shot noise limit (trace iii), respectively. To obtain the intensity difference squeezing, the probe and conjugate beam intensity noises are electronically subtracted, leading to a maximum intensity difference squeezing of $\sim$6~dB (trace iv, Fig.~\ref{fig:1}(b)) with respect to the shot noise limit (trace iii, Fig.~\ref{fig:1}(b)).

In our experiments, we estimate the total losses, including optical and detection losses, in each mode to be $\sim$22\%. With the given FWM gain of $\sim$11.5 in our experiment, we can calculate the individual probe and conjugate intensity noises from Eq.~(\ref{squeezing}) by setting either $\eta_a =0.78$ and $\eta_b =0$ (for the probe noise) or $\eta_a=0$ and $\eta_b =0.78$ (for the conjugate noise). With these settings, we obtain the intensity noise of the probe and conjugate $\sim$~12.4~dB above the shot noise level, which is in a close agreement with our experimental results. Furthermore, the observed intensity difference squeezing result of $\sim$6~dB is also consistent with the theoretical prediction from Eq.~(\ref{squeezing}). Therefore, we conclude that the intensity fluctuations in each mode are primarily of a quantum mechanical origin.
\begin{figure}[t]
    \centering
    \includegraphics[width=1\linewidth]{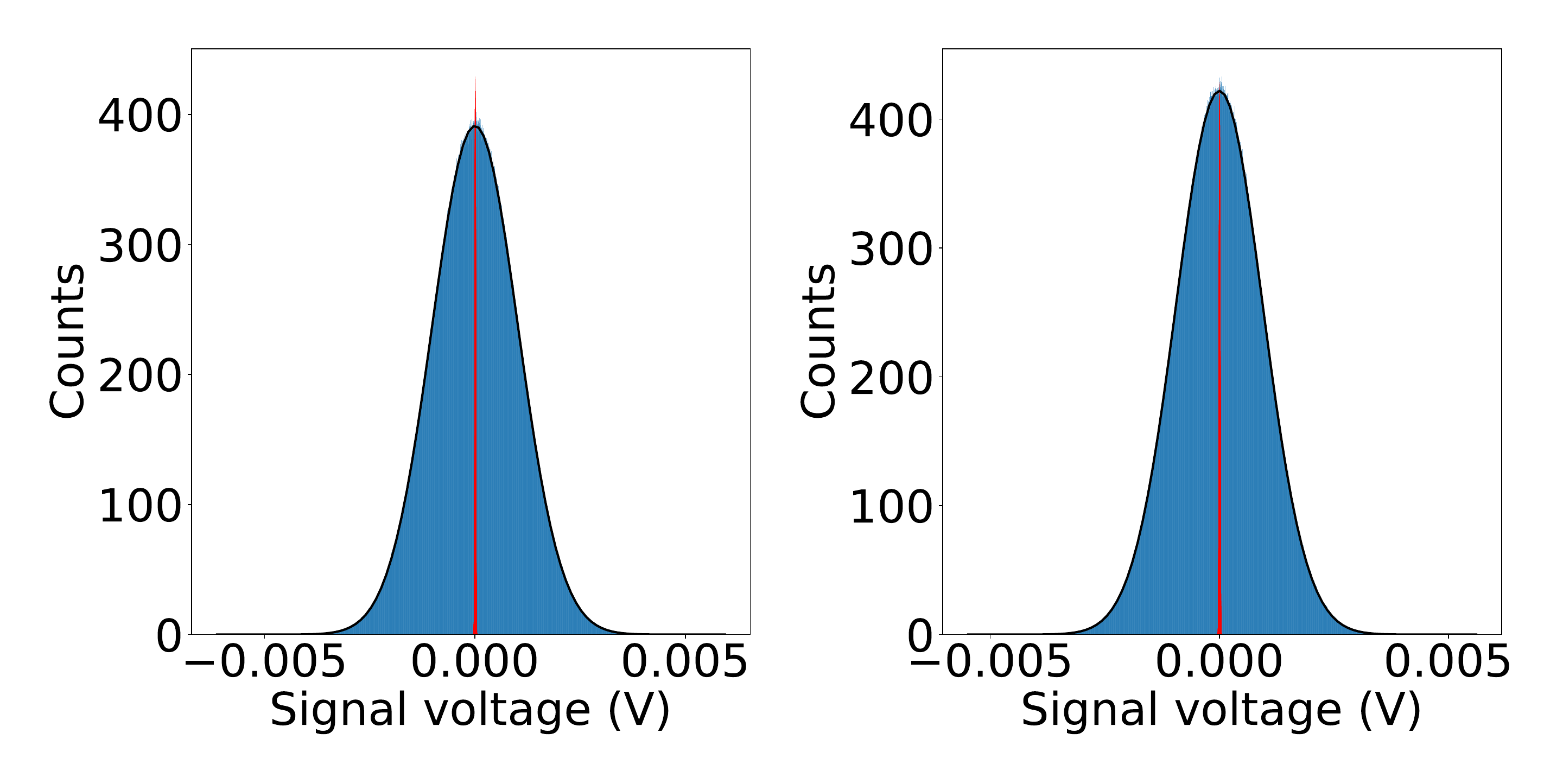}
    \caption{The intensity fluctuations histogram of the conjugate (left) and probe (right) fit to the normal distribution, with a sample size of $10^7$. The electronic noise histogram of the detector is also shown in the center (red).}
    \label{fig:2}
\end{figure}

To analyze the intensity fluctuations of the individual probe and conjugate beams over time, we use the I/Q waveform mode of the spectrum analyzers, which allows us to capture in-phase (I) and quadrature (Q) components of the RF signals. We simultaneously measure the I/Q waveform of the probe and conjugate intensity fluctuations by triggering both the spectrum analyzers with an external function generator at the same time. As a result, in Fig.~\ref{fig:1}(c), we show the same quadrature fluctuations in the probe and conjugate beams, represented by a voltage signal, over time. We use an analysis frequency of 2~MHz with 1~MHz of digital intermediate frequency (IF) bandwidth to record the intensity fluctuations. The digital IF bandwidth is the bandwidth of the low-pass filter applied after mixing with the signal at the chosen analysis frequency and fundamentally limits the rate of generation of random numbers\cite{shen2010practical}. As expected for the twin beams, we obtain highly correlated waveforms for the probe and conjugate. Furthermore, to quantify the amount of correlation, we calculated the normalized correlation coefficient between these waveforms as a function of the calculated time delay between the two data points, and the result is plotted in Fig.~\ref{fig:1}(d). It can be seen from the plot that a normalized correlation coefficient of more than 95\% is obtained between the two waveforms, which drops close to zero for even a single data delay.

In FIG. \ref{fig:2}, we have plotted the histograms of the probe and conjugate intensity fluctuations with a sample size of $10^7$. It can be seen that the histograms fit very well to a normal distribution. To extract the random numbers from these intensity fluctuations, we use the method of binning the Gaussian distributed data points into intervals containing an equal number of samples\cite{gabriel2010generator}, and assigning samples in each interval a sequence of \q{0}s and \q{1}s. 

The effective entropy due to quantum intensity fluctuations in each beam can be estimated as $H(X)_{eff} = H(X) - H(X)_c$, where $H(X)$ is the binary Shannon entropy\cite{gray2011entropy} of the measured intensity fluctuations and $H(X)_c$ is the same due to classical sources. In the present experiment, the classical contribution to the measured fluctuations comes from the electronic noise of the detection system. In the current analysis, to extract the strings of random numbers, we use $1.25\times10^8$ samples measured from the intensity fluctuations of each of the twin beams, and for the same sample size, the entropy of the electronic noise, i.e., $H(X)_c$, is also calculated to find the effective entropy $H(X)_{eff}.$ For $2^n$ intervals of binning, by definition, $H(X) = n$, i.e., we can extract $n$ bits of information.

\begin{figure}[t]
    \centering
    \includegraphics[width=1\linewidth]{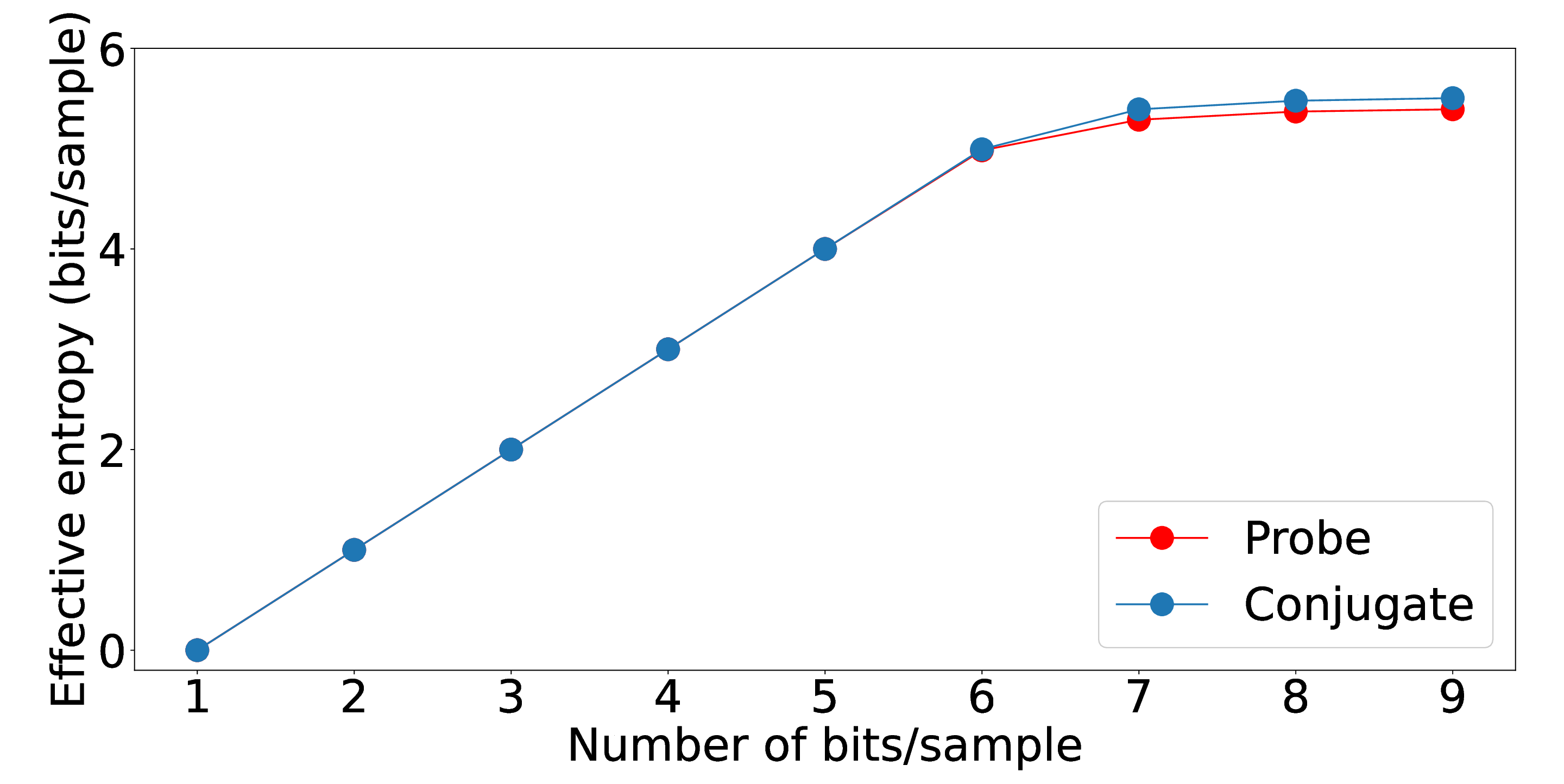}
    \caption{Effective entropy as a function of the number of bits extracted from the probe and conjugate intensity fluctuations.}
    \label{fig:3}
\end{figure}

In FIG.~\ref{fig:3}, we have plotted $H(X)_{eff}$ as a function of the number of bits extracted from the intensity fluctuations of the probe and conjugate. Clearly, the effective entropy starts to saturate after 8 bits per sample, or $2^8$ binning intervals, as the entropy of the classical noise starts to increase at the same rate as the entropy of the beam noise. Here, we obtain $H(X)_{eff}$ = 5.36 effective bits/sample for the probe and $H(X)_{eff}$ = 5.47 effective bits/sample for the conjugate when extracting 8 bits/sample from the intensity fluctuations.
Furthermore, to extract the quantum-correlated random numbers from the probe and conjugate intensity fluctuations, we extract 8 bits/sample from each waveform, and create a new sequence by manually selecting the common bits from both sequences. This can be thought of as the output of a classical reconciliation procedure in a QKD protocol. To eliminate the influence of the classical sources of noise in the extracted bits, we perform hashing of the bits to produce a final sequence that is of very high statistical quality and that has purely quantum origin. This is a direct consequence of the leftover hash lemma\cite{tomamichel2011leftover}. We implement the hashing using the SHA512 hash function\cite{gabriel2010generator} in Python, with the hashed string effectively containing 3.2 bits per sample of the original waveforms and extract a sequence of random numbers. 

\begin{figure}[t]
    \centering
    \includegraphics[width=1\linewidth]{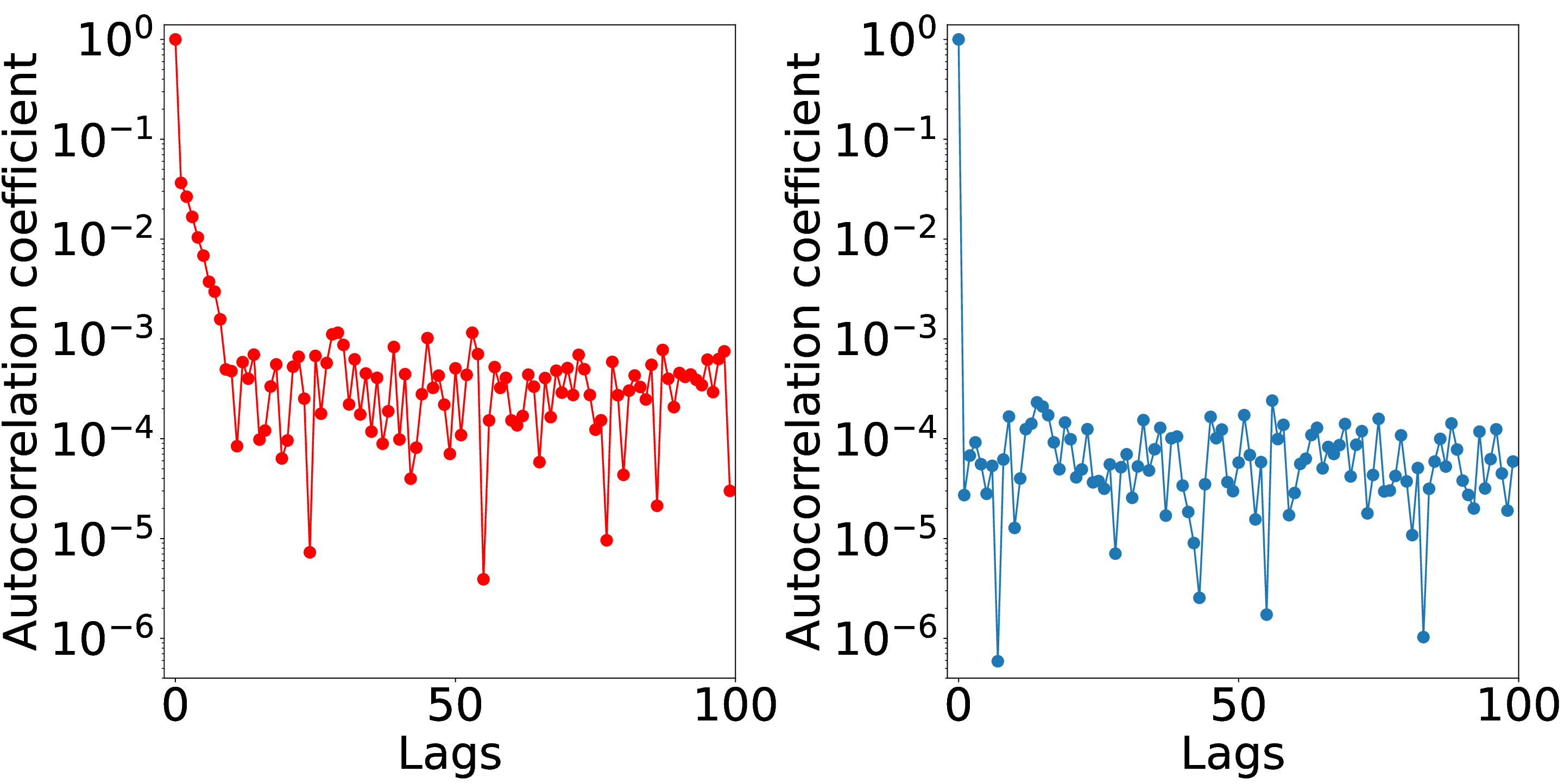}
    \caption{Autocorrelation coefficients (absolute value) of the raw (left) and hashed sequences (right), as a function of lags of each sequence. $10^8$ bits are used to compute the autocorrelation coefficients.}
    \label{fig:4}
\end{figure}
\begin{table}[b]
    \centering
    \begin{tabular}{ccc} 
            \hline Statistical&Proportion passed&~~~~P - value\\
                    Test&after hashing&~~~~\\
            \hline Frequency&0.985& 0.59\\ 
            BlockFrequency&0.9775& 0.41\\ 
            CumulativeSums&0.985& 0.41\\ 
            Runs& 0.9825& 0.53\\ 
            LongestRun& 0.995&0.81\\ 
            Rank& 0.99& 0.77\\ 
            FFT& 0.9925& 0.14\\
            NonOverlappingTemplate& 0.975& 0.04\\ 
            OverlappingTemplate&0.9925& 0.54\\ 
            Universal&0.995& 0.7\\ 
            ApproximateEntropy&0.9975& 0.09\\
            RandomExcursions& 0.979 & 0.58\\
            RandomExcursionsVariant& 0.979& 0.62\\
            Serial& 0.9875 & 0.02\\
            LinearComplexity& 0.9875& 0.92\\
    \end{tabular}
    \caption{The results of the statistical tests from the NIST suite after hashing. Some tests have many variants for which the worst proportion is shown. Each test outputs a global p-value using a goodness-of-fit test for the 400 sequences that have been tested. The p-values of all tests for the hashed variant lie above the expected value of 0.0001\cite{bassham2010sp}.}
    \label{tab:1}
\end{table}

To perform the statistical tests, we feed 400 sequences, each containing 1 million bits, to the NIST statistical suite\cite{bassham2010sp}. The proportion of hashed sequences passing each test is shown in TABLE \ref{tab:1}. For our sample size, the proportion of sequences passing each test should lie above\cite{bassham2010sp} 0.975 for all tests except the random excursion and above 0.97 for the random excursion tests. For tests providing multiple proportions, the worst case is shown. The proportions passing the tests clearly show that the final sequence is of good statistical quality. The p-values of the hashed sequences are also shown. The hashed sequence also passes the Rabbit battery of tests from the TestU01 suite\cite{l2007testu01}. 

The autocorrelation function (see FIG. \ref{fig:4}) of the hashed bits also exhibits a near zero value for all non-zero lags, indicating that negligible information about the next or previous bits can be extracted with knowledge of any given bit.

The bit generation rate of quantum-correlated random numbers in our system is $\sim$6 Mbps, assuming the post-processing time is negligible compared to the data acquisition time. This rate can be further enhanced by increasing the squeezing bandwidth, which in turn expands the detection bandwidth. While the current work presents a proof-of-principle experiment to create a sequence of quantum random numbers with 6~dB of the intensity difference squeezing, a similar system\cite{liu2019interference} having intensity difference squeezing of over 10 dB would significantly enhance the final bit rate of the quantum random number generator through a dual mechanism. Firstly, a higher level of squeezing contributes to increased intrinsic quantum entropy within the raw signal. Even as the noise difference between the twin beams is reduced, the individual mode intensity noise increases, providing a richer source of inherent randomness. Secondly, stronger squeezing leads to improved correlations between the twin beams. This enhanced correlation is crucial for the post-selection process, as it allows for a greater proportion of "common bits" to be extracted from the raw data. Consequently, less data needs to be discarded, leading to a more efficient and higher-rate generation of random numbers.

The quantum random numbers generated in the current work operates within a trusted-device framework\cite{ma2016quantum}, where the security relies on the precise characterization and assumed integrity of the physical components, particularly the twin-beam source and detection system. While this design necessitates trust in the device's adherence to its physical model, we account for an adversary with knowledge of device properties and classical noise through robust post-processing guided by the Leftover Hash Lemma\cite{tomamichel2011leftover}. This ensures the security of the final random bit string even if the raw data contains classical correlations, effectively transforming a high-entropy quantum input into a uniformly distributed, adversary-unknown output. This approach offers a pragmatic balance, enabling high random number generation rates crucial for many applications, distinguishing it from device-independent schemes that, while offering ultimate security guarantees through fundamental physical inequalities, typically incur higher experimental complexity and significantly lower rates\cite{christensen2013detection}.

The scheme of generation of correlated quantum random number sequences using bright twin beams from the four-wave mixing offers significant advantages for practical quantum communication applications. Specifically, these sequences are ideal for continuous-variable quantum key distribution and enable the high-speed distribution of one-time pads in symmetric cryptography\cite{grosshans2002continuous,silberhorn2002quantum,pirandola2008continuous,gehring2015implementation,madsen2012continuous}. This directly facilitates the fast and secure distribution of cryptographic keys, substantially enhancing the practical relevance of QRNGs in quantum information science.

\section*{AUTHOR DECLARATIONS}
\subsection*{Conflict of Interest}
The authors have no conflicts to disclose
\subsection*{Author Contributions}
\textbf{Anirudh Shekar:} Data curation (equal); formal analysis (lead); investigation (equal); methodology (lead); software (lead); validation (equal); visualization (lead); writing - original draft (lead); writing - review and editing (equal). \textbf{Chirang R. Patel:} Investigation (equal); data curation (equal); validation (equal); writing - review and editing (equal). \textbf{Jerin A. Thachil:} Investigation (equal); data curation (equal); validation (equal); writing - review and editing (equal). \textbf{Ashok Kumar:} Conceptualization (lead); formal analysis (supporting); project administration (lead); resources (lead); supervision (lead); validation (equal); writing - original draft (supporting); writing - review and editing (equal).
\section*{Data Availability Statement}
The data that support the findings of this study are available from the corresponding author upon reasonable request.
\bibliography{aipsamp}

\end{document}